\documentclass[aps,prl,showpacs,floatfix,twocolumn,byrevtex]{revtex4-1}
\usepackage{amsmath}
\usepackage{amssymb}
\usepackage{amstext}
\usepackage{amsopn}
\usepackage{amsfonts}
\usepackage{amsxtra}
\usepackage[english]{babel}
\usepackage{graphicx}
\usepackage{float}
\usepackage{bm}
\usepackage{multirow}
\usepackage{dcolumn}
\usepackage{color}
\usepackage{hyperref}

\newcommand{\icm}{\ensuremath{\textrm{cm}^{-1}}}

\newcommand{\BSTS}{Sn-BSTS}

\begin{document}

\title{Van Hove Singularity Arising from Mexican-Hat-Shaped Inverted Bands in the Topological Insulator Sn-doped Bi$_{1.1}$Sb$_{0.9}$Te$_{2}$S}
\author{Wenchao Jiang}
\author{Bowen Li}
\author{Xiaomeng Wang}
\author{Guanyu Chen}
\author{Tong Chen}
\author{Ying Xiang}
\author{Wei Xie}
\author{Yaomin Dai}
\email{ymdai@nju.edu.cn}
\author{Xiyu Zhu}
\author{Huan Yang}
\author{Jian Sun}
\email{jiansun@nju.edu.cn}
\author{Hai-Hu Wen}
\email{hhwen@nju.edu.cn}
\affiliation{National Laboratory of Solid State Microstructures and Department of Physics, Collaborative Innovation Center of Advanced Microstructures, Nanjing University, Nanjing 210093, China}

\date{\today}
%
%

\begin{abstract}
The optical properties of Sn-doped Bi$_{1.1}$Sb$_{0.9}$Te$_{2}$S, the most bulk-insulating topological insulator thus far, have been examined at different temperatures over a broad frequency range. No Drude response is detected in the low-frequency range down to 30~cm$^{-1}$, corroborating the excellent bulk-insulating property of this material. Intriguingly, we observe a sharp peak at about 2\,200~cm$^{-1}$ in the optical conductivity at 5~K. Further quantitative analyses of the line shape and temperature dependence of this sharp peak, in combination with first-principles calculations, suggest that it corresponds to a van Hove singularity arising from Mexican-hat-shaped inverted bands. Such a van Hove singularity is a pivotal ingredient of various strongly correlated phases.
\end{abstract}



\maketitle

%
%
Stimulated by the recent discovery of unconventional superconductivity (SC)~\cite{Cao2018NatureSC} and correlated insulating behavior~\cite{Cao2018NatureMott} in `magic-angle' twisted bilayer graphene, flat bands have quickly become one of the hottest topics in condensed matter physics~\cite{Marchenko2018SA,Yankowitz2019Science,Yin2019NP,Li2019PRL,Tarnopolsky2019PRL}. Electronic systems having flat bands close to the Fermi level ($E_F$) are of great interest, because in such systems, with the quantum kinetic energy being quenched, the ground state is achieved by minimizing Coulomb interaction energy, which consequently results in a variety of strongly correlated phases, such as unconventional SC~\cite{Cao2018NatureSC}, Mott-like insulator~\cite{Cao2018NatureMott}, and Wigner crystallization~\cite{Wu2007PRL}. In addition, flat bands give rise to a van Hove singularity (VHS) in the density of states (DOS). Theoretical studies have shown that such a VHS, when situated near $E_F$, can induce exotic quantum phenomena including the fractional quantum Hall effect~\cite{Sun2011PRL,Tang2011PRL,Neupert2011PRL}, charge density wave (CDW)~\cite{Rice1975PRL}, magnetism~\cite{Zhang2010PRA,Tasaki1998PTP,Okada2001PRL} and SC~\cite{Hirsch1986PRL,Newns1992PRL,Tang2014NP,Kopnin2011PRB}.

Owing to the fascinating physics presented by flat-band systems, the search for materials with flat bands near $E_F$ is highly motivated. To date, flat bands have been realized in several systems including twisted bilayer graphene~\cite{Cao2018NatureSC,Cao2018NatureMott,Marchenko2018SA,Li2010NP}, high-$T_c$ cuprates~\cite{Gofron1994PRL,Dessau1993PRL}, kagome magnet Co$_{3}$Sn$_{2}$S$_{2}$~\cite{Yin2019NP}, heavy fermion compounds~\cite{Hoshino2013PRL,Pfau2013PRL}, as well as some lattice models~\cite{Wu2007PRL,Sun2011PRL,Tang2011PRL,Neupert2011PRL,Zhang2010PRA}; the exotic quantum phenomena found in these systems are believed to be closely tied to the flat bands. For instance, in `magic-angle' twisted bilayer graphene, the electronic bands near $E_{F}$ are flat, and VHSs associated with these flat bands have been experimentally detected~\cite{Li2010NP}. Mott-like insulating states are discovered at half-filling of these flat bands~\cite{Cao2018NatureMott}, while unconventional SC emerges upon doping the material away from the correlated insulating states~\cite{Cao2018NatureSC}. In the high-$T_c$ cuprate YBa$_{2}$Cu$_{4}$O$_{8}$~\cite{Gofron1994PRL}, an ``extended'' VHS has been observed at about 19~meV below $E_F$. This VHS, which arises from a flat band along the $\Gamma$-$Y$ direction, is believed to be responsible for the high SC transition temperature. Recently, Yin \emph{et al}.~\cite{Yin2019NP} reveal a VHS at 8~meV below $E_F$ stemming from a flat band with non-trivial Berry phase in the kagome magnet Co$_{3}$Sn$_{2}$S$_{2}$, which is likely to be the origin of the negative magnetism they observed. Theoretical calculations based on some lattice models~\cite{Sun2011PRL,Tang2011PRL,Neupert2011PRL} have revealed flat bands with a nonzero Chern number; partial filling of these flat bands can lead to fractional quantum Hall states without magnetic field.

\begin{figure}[tb]
\includegraphics[width=\columnwidth]{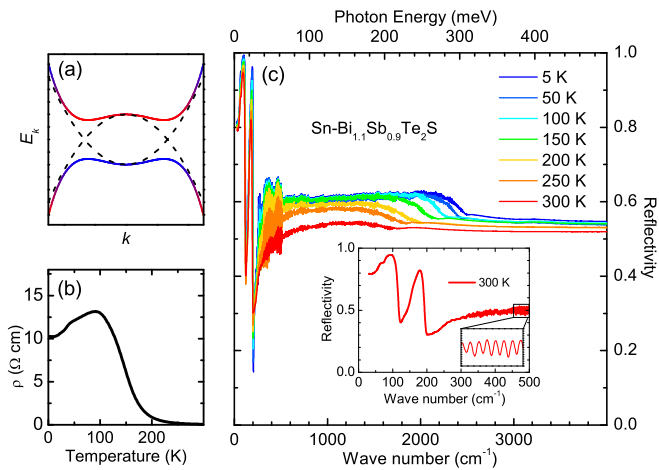}
\caption{(a) Schematic plot of the Mexican-hat dispersion (solid lines) formed by deeply inverted bands (dashed lines) in TIs. (b) $T$-dependent resistivity of \BSTS. (c) Reflectivity of \BSTS\ up to 4\,000~\icm\ at different temperatures. The inset shows the far-infrared reflectivity at 300~K.}
\label{VHSRef}
\end{figure}

In addition to these systems, topological insulators (TIs) are also expected to possess flat bands in the bulk. Theoretical calculations have pointed out that deeply inverted bands in TIs unavoidably develop a ``Mexican-hat-shaped'' dispersion~\cite{Yazyev2012PRB,He2014arXiv}, as shown by the schematic in Fig.~\ref{VHSRef}(a). Such Mexican-hat-shaped bands give rise to a VHS in the DOS at the band edge, which may not only account for the symmetry breaking instability of TIs~\cite{He2014arXiv}, such as magnetism~\cite{Zhang2013Science} and SC~\cite{Hor2010PRL,Liu2015JACS,Zhong2014PRB,An2018PRB,Zhang2011PNAS,Cai2018PRM,Cui2019CPL}, but also provide a platform for exploring exotic quantum phenomena, particularly the topological SC~\cite{Fu2008PRL,Fu2010PRL}.

Optical spectroscopy is a powerful technique to reveal such a VHS in TIs if $E_F$ lies in the bulk band gap, because electronic transitions between Mexican-hat-shaped bands inevitably produce a divergently sharp peak in the optical conductivity, which measures the joint density of states (JDOS)~\cite{Dressel2002,Basov2005RMP}. Optical studies of TIs have mainly focused on the band gap and bulk carriers~\cite{Akrap2012PRB,Pietro2012PRB,Aleshchenko2014JETPL,Dubroka2017PRB,LaForge2010PRB,Post2013PRB,Xi2014PRL}, the phonon modes~\cite{Akrap2012PRB,LaForge2010PRB}, as well as the topological surface states~\cite{Reijnders2014PRB,Post2015PRL}. We notice that an absorption peak near the bulk band gap appears in the optical conductivity of Bi$_{2}$Te$_{2}$Se~\cite{Akrap2012PRB,Pietro2012PRB,Dubroka2017PRB,Reijnders2014PRB,Aleshchenko2014JETPL}. However, no detailed analysis of this peak has been done. To date, evidence for the VHS arising from Mexican-hat-shaped inverted bands in TIs remains elusive.

Thus far, Sn-doped Bi$_{1.1}$Sb$_{0.9}$Te$_{2}$S (\BSTS) has been found to be the most bulk-insulating TI whose $E_F$ sits in the bulk band gap~\cite{Kushwaha2016NC}, providing a perfect system to trace the VHS arising from Mexican-hat-shaped inverted bands through optical spectroscopy. The only optical study on this material is in the THz range, focusing on the topological surface states~\cite{Cheng2016PRB}. Here, we present a detailed optical study on \BSTS. Our optical conductivity shows no Drude response down to 30~cm$^{-1}$, confirming the bulk-insulating nature of this material. A sharp peak is clearly observed at about 2\,200~\icm\ in the optical conductivity at 5~K. By comparing our experimental results with theoretically calculated optical conductivity and band structure, we conclude that the sharp peak in the optical conductivity represents a VHS associated with Mexican-hat-shaped inverted bands.

%
%
High-quality Bi$_{1.08}$Sn$_{0.02}$Sb$_{0.9}$Te$_{2}$S single crystals with large and shiny surfaces were synthesized using the Bridgeman method~\cite{Kushwaha2016NC}. Figure~\ref{VHSRef}(b) shows the resistivity $\rho$ of our sample as a function of temperature ($T$), which agrees well with published data~\cite{Kushwaha2016NC}. In the high-temperature region, $\rho$ increases exponentially with decreasing $T$, suggesting that the transport properties are dominated by the insulating bulk states. The decrease of $\rho$ at low $T$s is a signature of surface-dominated transport~\cite{Kushwaha2016NC,Ren2010PRB,Taskin2011PRL}: the bulk resistivity becomes so large at low $T$ that it is short-circuited by the metallic topological surface states. The details about our optical measurements and Kramers-Kronig analysis are given in the supplementary material~\cite{SuppMat} (see, also, ref.~\cite{Homes1993AO}).

%

%
%
Figure~\ref{VHSRef}(c) displays the measured reflectivity $R(\omega)$ up to 4\,000~\icm\ at several selected temperatures; the inset shows $R(\omega)$ in the far-infrared range at 300~K, which is dominated by strong IR-active phonons~\cite{LaForge2010PRB,Akrap2012PRB}. In the presence of free carriers, $R(\omega)$ would exhibit an upturn in the low-frequency range~\cite{Akrap2012PRB,Reijnders2014PRB}. Such an upturn is not observed in the low-frequency $R(\omega)$ of \BSTS\ as shown in the inset of Fig.~\ref{VHSRef}(c), indicative of an insulating sample. The most striking feature in $R(\omega)$ at 5~K is a step at 2\,400~\icm. This step feature diminishes and shifts to lower frequency as $T$ rises. Moreover, fringes resulting from interference between light reflected from the top and bottom surfaces appear below the step, and disappear at about 250~\icm\ where phonon absorption emerges. The fringes can be seen more clearly in the zoomed-in window in the inset of Fig.~\ref{VHSRef}(c). The emergence of interference fringes suggests that the sample becomes transparent in this frequency range.

\begin{figure}[tb]
\includegraphics[width=\columnwidth]{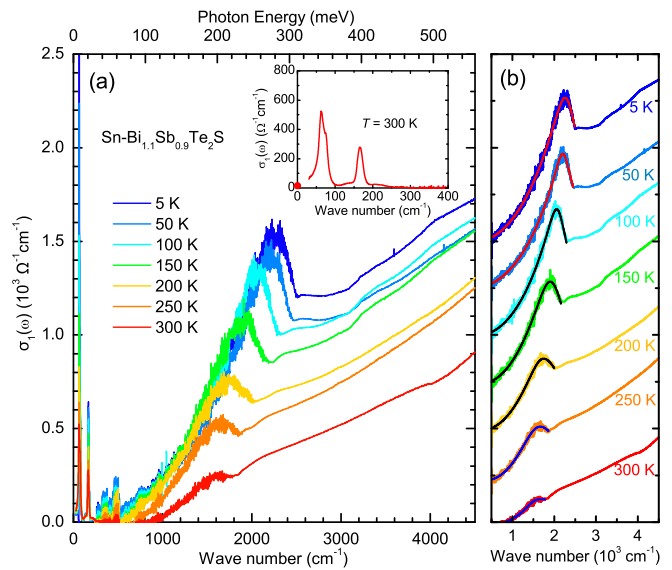}
\caption{(a) Optical conductivity of \BSTS\ up to 4\,500~\icm\ at several representative temperatures. Inset: Far-infrared optical conductivity at 300~K, showing IR-active phonons. The red solid circle at zero frequency denotes the transport value of the dc conductivity at 300~K. (b) Optical conductivity with offset, highlighting the sharp peak between 1\,600 and 2\,200~\icm. Solid lines through the data are fitting results at different temperatures.}
\label{VHSS1}
\end{figure}

The real part of the optical conductivity $\sigma_{1}(\omega)$ provides more straightforward information, as it is directly related to the JDOS~\cite{SuppMat}. Figure~\ref{VHSS1}(a) shows $\sigma_{1}(\omega)$ of \BSTS\ at different $T$s up to 4\,500~\icm. The inset depicts $\sigma_{1}(\omega)$ measured at 300~K below 400~\icm, where IR-active phonons can be clearly identified as discrete sharp peaks. The red solid circle at zero frequency represents the dc conductivity at 300~K from transport measurements, which agrees well with the low-frequency extrapolation of $\sigma_{1}(\omega)$. In the low-frequency region, $\sigma_{1}(\omega)$ is vanishingly small, and no Drude response is evidently resolved for all temperatures, consistent with the insulating nature of our sample. For $\sigma_{1}(\omega)$ at 5~K, with increasing frequency, an absorption edge at $\sim$1\,500~\icm\ develops, leading to a sharp peak at $\sim$2\,200~\icm. As $T$ is elevated, this peak becomes weaker and quickly shifts to lower frequency. The noise-like signal setting in below the peak is interference fringes as discussed for $R(\omega)$.

In the following, we concentrate on the sharp peak at $\sim$2\,200~\icm\ in the $\sigma_{1}(\omega)$ spectra. Since IR-active phonons are not expected at such a high frequency, we deduce that this peak originates from electronic transitions. Furthermore, we can safely rule out electronic transitions associated with impurity bands as the origin of such a sharp peak in $\sigma_{1}(\omega)$ of Sn-BSTS~\cite{SuppMat,Gaymann1995PRB}. As $\sigma_{1}(\omega)$ is proportional to the JDOS, a quasidivergent peak in $\sigma_{1}(\omega)$ explicitly signifies a VHS in the JDOS of \BSTS~\cite{Hove1953PR}.

Further information about the VHS in \BSTS\ may be obtained through a quantitative analysis of $\sigma_{1}(\omega)$. Van Hove has pointed out that the DOS of electron bands has singularities at points where $|\nabla_k(E)|$ vanishes~\cite{Hove1953PR}. These points in the band structure are known as critical points and the corresponding singularities in the DOS are called VHSs. The optical response of interband critical points can be described by a standard analytic line shape for the complex dielectric function~\cite{Lautenschlager1987PRB}
%
\begin{equation}
\tilde{\varepsilon}(\omega) = C - A e^{i\phi} (\omega - E_{t} + i \Gamma)^{n},
\label{EqCP}
\end{equation}
where $A$, $E_t$, $\Gamma$ and $\phi$ stand for the amplitude, energy threshold, broadening and excitonic phase angle of the critical point, respectively. For three-dimensional (3D) critical points $n = 1/2$. The real part of the optical conductivity $\sigma_{1}(\omega)$ and the imaginary part of the dielectric function $\varepsilon_{2}(\omega)$ have the following simple relation:
%
\begin{equation}
\sigma_{1}(\omega)=\frac{2\pi}{Z_0}\omega \varepsilon_{2}(\omega),
\label{S1E2}
\end{equation}
where $Z_{0}$ = 377~$\Omega$ is the vacuum impedance. We fit this equation to the measured $\sigma_{1}(\omega)$ at all $T$s and plot the fitting results as solid lines through the data in Fig.~\ref{VHSS1}(b). Note that Eq.~(\ref{EqCP}) gives an excellent description to the line shape of $\sigma_{1}(\omega)$ at all measured temperatures. From the fitting, we can extract the $T$ dependence of two key parameters: $E_t$ the energy threshold of the VHS, and $\phi$ which contains information about the type of the VHS or the dispersion of the electronic bands.

\begin{figure}[tb]
\includegraphics[width=\columnwidth]{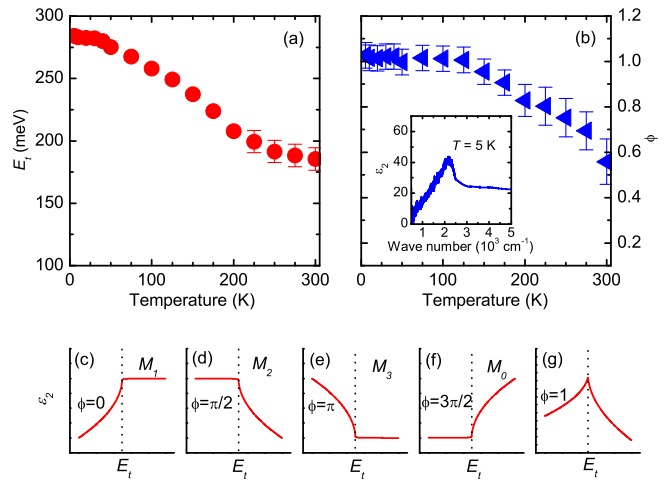}
\caption{$T$ dependence of $E_t$ (a) and $\phi$ (b) extracted from the fit at all measured temperatures. The inset of (b) displays $\varepsilon_{2}$ of Sn-BSTS measured at 5~K. (c)-(f) Schematic diagrams of $\varepsilon_{2}(\omega)$ for different types of VHSs. (g) Schematic plot of $\varepsilon_{2}(\omega)$ for a VHS with $\phi = 1$.}
\label{VHSTD}
\end{figure}

As portrayed by Fig.~\ref{VHSTD}(a), $E_{t}$ increases with decreasing $T$. This implies that the bulk band gap increases as $T$ is lowered. In TIs, inverted and non-inverted band gaps exhibit opposite temperature dependence~\cite{Strauss1967PR,Yazyev2012PRB}. Studies on TIs with a small band gap, such as Pb$_{1-x}$Sn$_{x}$Te~\cite{Dimmock1966PRL}, Pb$_{1-x}$Sn$_{x}$Se~\cite{Strauss1967PR,Dziawa2012NM} and ZrTe$_{5}$~\cite{Xu2018PRL}, have revealed that non-inverted gaps decrease upon cooling, while gaps between inverted bands increase with decreasing $T$~\cite{Dimmock1966PRL,Strauss1967PR,Dziawa2012NM,Xu2018PRL}. Based on these properties of TIs, it is tempting to associate the VHS in \BSTS\ with inverted bands. However, the increase of the band gap with decreasing $T$ may not be taken as conclusive evidence for inverted bands, since the band gap in conventional semiconductors, e.g. GaAs~\cite{Lautenschlager1987PRB}, Si and Ge~\cite{Lautenschlager1985PRB}, also increase with decreasing $T$. Further evidence for the VHS associated with inverted bands in \BSTS\ will be provided by theoretical calculations, and a comparison between the calculations and experimental data.

Having examined the $T$ dependence of $E_t$, we next look into $\phi$ and its evolution with $T$. Figure~\ref{VHSTD}(b) shows that $\phi \simeq 1$ at low temperatures, and it decreases with increasing $T$. For a 3D system, there are four types of VHSs corresponding to four types of interband critical points (labeled $M_{1}$, $M_{2}$, $M_{3}$, and $M_{0}$) in the band structure~\cite{Hove1953PR,Lautenschlager1987PRB,Yu2010}; $\phi$ takes the values of 0, $\pi/2$, $\pi$, and $3\pi/2$ for $M_{1}$, $M_{2}$, $M_{3}$, and $M_{0}$, respectively. Figures~\ref{VHSTD}(c)-\ref{VHSTD}(f) schematically illustrate $\varepsilon_{2}(\omega)$ calculated from Eq.~(\ref{EqCP}) for all four types of VHSs. While $M_0$ [Fig.~\ref{VHSTD}(f)] and $M_3$ [Fig.~\ref{VHSTD}(e)] refer to a minimum and a maximum in the interband separation, respectively, $M_1$ and $M_2$ represent saddle points [Figs.~\ref{VHSTD}(c) and \ref{VHSTD}(d)]. The experimentally determined $\phi$ for Sn-BSTS does not have a value of integer multiples of $\pi/2$. This implies that the VHS in Sn-BSTS does not simply fall into the basic classification of VHSs $M_{0}$-$M_{3}$. In addition, $\phi$ varies with $T$, from which we may deduce that the electronic band dispersion changes with $T$.

It's worthwhile to notice that regular VHSs manifest themselves as step-like features in $\varepsilon_{2}(\omega)$, whereas the one with $\phi = 1$, as shown in Fig.~\ref{VHSTD}(g), produces a divergently sharp peak. Such a peak, which can also be clearly identified in the measured $\varepsilon_{2}(\omega)$ [inset of Fig.~\ref{VHSTD}(b)], unequivocally suggests that the VHS in Sn-BSTS reflects a more singular JDOS than regular ones. This recalls the expected Mexican-hat-shaped dispersion for deeply inverted bands in TIs~\cite{He2014arXiv,Yazyev2012PRB}. Theoretical calculations based on an analytical model~\cite{He2014arXiv} have shown that for non-inverted and shallowly inverted bands, a step feature, representing a regular VHS, is obtained in the DOS around the band edge. In stark contrast, for deeply inverted bands, a sharp peak, which corresponds to a VHS arising from Mexican-hat-shaped bands, emerges in the DOS around the band edge. These results point to that the VHS in Sn-BSTS is most likely associated with Mexican-hat-shaped inverted bands.

In order to provide more insights into the VHS as well as the electronic band dispersion in \BSTS, we performed first-principles calculations~\cite{SuppMat,Abt1994PB,Ambrosch-Draxl2006CPC}.
\begin{figure}[tb]
\includegraphics[width=\columnwidth]{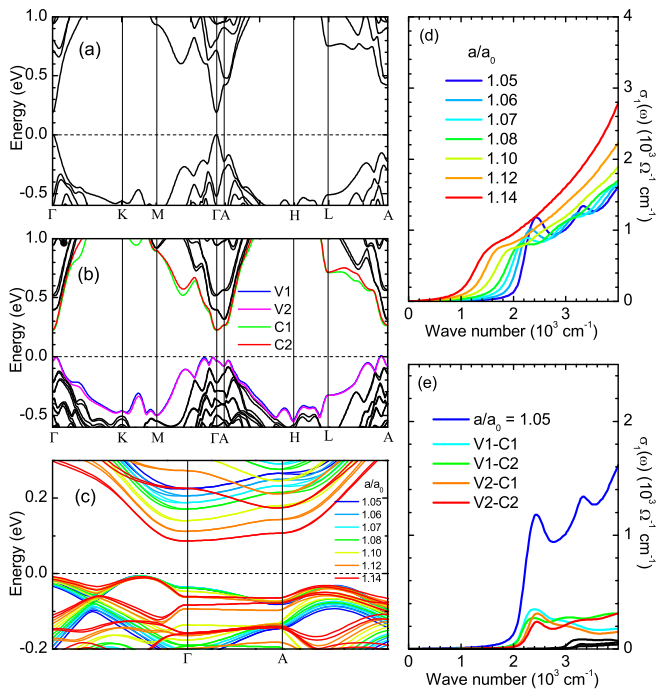}
\caption{Calculated band structures of BiSbTe$_{2}$S at $a/a_{0}=1.05$ (a) without SOC and (b) with SOC. (c) Calculated band structure at different $a/a_{0}$, simulating the effect of $T$ on the band structure. (d) Calculated $\sigma_{1}(\omega)$ at various $a/a_{0}$. (e) Calculated $\sigma_{1}(\omega)$ at $a/a_{0} = 1.05$ (blue curve) which is decomposed into contributions from different band combinations.}
\label{VHSCal}
\end{figure}
Figures~\ref{VHSCal}(a) and \ref{VHSCal}(b) compare the calculated band structures of BiSbTe$_{2}$S at $a/a_{0}=1.05$ without and with SOC. A conspicuous change in the band structure induced by turning on SOC is an anti-crossing feature near the $\Gamma$ point, which is a prototypical signature of inverted bands. The band inversion can also be realized from the calculated orbit-resolved band structure~\cite{SuppMat}. Figure~\ref{VHSCal}(c) shows an enlarged view of the calculated band structure near the $\Gamma$ point for various $a/a_{0}$ ratios. It is noticeable that the inverted bands exhibit distinct Mexican-hat-shaped dispersion, which is expected to yield a VHS in the JDOS~\cite{He2014arXiv}. The separation between the Mexican-hat-shaped inverted bands increases as $a/a_{0}$ is reduced, in accord with the $T$ dependence of $E_{t}$ determined from the measured $\sigma_{1}(\omega)$. Furthermore, the dispersion of these inverted bands varies slightly with $a/a_{0}$. Such an effect has been inferred from the change of $\phi$ with $T$ extracted from $\sigma_{1}(\omega)$. All these facts explicitly indicate that the VHS we observed in $\sigma_{1}(\omega)$ arises from the Mexican-hat-shaped inverted bands near the $\Gamma$ point.

To further support our conclusion, we calculated $\sigma_{1}(\omega)$ at different $a/a_{0}$, as well as contributions from different band combinations to $\sigma_{1}(\omega)$ for $a/a_{0} = 1.05$. The calculated $\sigma_{1}(\omega)$ at $a/a_{0} = 1.05$, shown as the blue curve in Fig.~\ref{VHSCal}(d), agrees well with the measured $\sigma_{1}(\omega)$ at 5~K. A sharp peak can be clearly seen at about 2\,400~\icm. Such a sharp peak is absent in $\sigma_{1}(\omega)$ calculated from the band structure without SOC~\cite{SuppMat}, indicating its close relation with inverted bands. With increasing $a/a_{0}$, the peak in the calculated $\sigma_{1}(\omega)$ diminishes and shifts towards lower frequency, qualitatively mimicking the $T$ dependence of the measured $\sigma_{1}(\omega)$. In Fig.~\ref{VHSCal}(e), the calculated $\sigma_{1}(\omega)$ at $a/a_{0} = 1.05$ (blue curve) is decomposed into contributions from different band combinations. More specifically, the cyan, green, orange and red curves in Fig.~\ref{VHSCal}(e) denote the $\sigma_{1}(\omega)$ components contributed by the electronic transitions V1-C1, V1-C2, V2-C1 and V2-C2, respectively. Here, the electronic bands V1, V2, C1 and C2 are traced out as colored lines in Fig.~\ref{VHSCal}(b). The black curves in Fig.~\ref{VHSCal}(e) represent contributions from electronic transitions between bands at higher energies [black lines in Fig.~\ref{VHSCal}(b)]. It is obvious from Fig.~\ref{VHSCal}(e) that only electronic transitions between the inverted bands near the $\Gamma$ point contribute to the sharp peak at about 2\,400~\icm\ in $\sigma_{1}(\omega)$. This offers further support for our conclusion that the sharp peak we observed in the $\sigma_{1}(\omega)$ of \BSTS\ corresponds to a VHS induced by the deeply inverted bands with Mexican-hat dispersions near the $\Gamma$ point.

Finally, we notice that the calculated $\sigma_{1}(\omega)$ [Fig.~\ref{VHSCal}(e)] can not describe the presence of absorption below the VHS, which gives rise to a much broader absorption edge in the measured $\sigma_{1}(\omega)$ [Fig.~\ref{VHSS1}(a)]. A close inspection of the band structures in Figs.~\ref{VHSCal}(b) and \ref{VHSCal}(c) reveals that the distortions of the valence and conduction bands towards a Mexican-hat shape are asymmetric, resulting in an indirect band gap of $\sim$230~meV near the $\Gamma$ point. The energy of this indirect band gap is much lower than the VHS energy threshold $\sim$2\,400~\icm\ (297~meV) in the calculated $\sigma_{1}(\omega)$. This means that indirect transitions can occur below the VHS energy threshold. Since indirect electronic transitions are not included in our first-principles calculations, it is natural that no low-energy absorption is present in the calculated $\sigma_{1}(\omega)$. In contrast, electronic transitions across the indirect band gap should have considerable contributions to the experimental $\sigma_{1}(\omega)$, which may account for the presence of absorption below the VHS, i.e. a much broader absorption onset in the measured $\sigma_{1}(\omega)$.

%
The observation of a VHS arising from Mexican-hat-shaped inverted bands in a TI deserves further attention, because it may not only elucidate the origin of many observed symmetry breaking phases in TIs, but also pave the way for the realization of more exotic quantum phenomena. Extensive studies have shown that systems with a VHS in the DOS resulting from flat bands near $E_{F}$ are generally unstable against symmetry breaking phases, such as CDW~\cite{Rice1975PRL}, SC~\cite{Hirsch1986PRL,Newns1992PRL,Tang2014NP,Kopnin2011PRB} and magnetism~\cite{Zhang2010PRA,Tasaki1998PTP,Okada2001PRL,Yin2019NP}. In this sense, the magnetic quantum phase transition in Cr-doped Bi$_{2}$(Se$_{x}$Te$_{1-x}$)$_{3}$ that is driven by bulk band topology~\cite{Zhang2013Science}, and the emergence of SC induced by doping~\cite{Hor2010PRL,Liu2015JACS,Zhong2014PRB}, pressure~\cite{An2018PRB,Zhang2011PNAS,Cai2018PRM} or protonation~\cite{Cui2019CPL} in TIs are likely to be intimately tied to the VHS resulting from Mexican-hat-shaped inverted bands. Moreover, driving a TI into the superconducting state may lead to a topological superconductor~\cite{Fu2008PRL,Fu2010PRL}, whose surface states host Majorana fermions which have potential applications in topological quantum computing. It is interesting that \BSTS\ exhibits pressure-induced SC with the maximum $T_c \simeq 12$~K (the highest among pressurized TIs), but structural phase transitions have also been observed~\cite{An2018PRB,Cai2018PRM}, which does not guarantee the topological states of this material. Nevertheless, with such an unusual VHS, \BSTS\ may be a promising system for the realization of topological SC, if the Mexican-hat-shaped inverted bands are tuned to $E_F$ through doping~\cite{Hor2010PRL,Liu2015JACS,Zhong2014PRB}, ionic liquid gating~\cite{Lei2016PRL} or protonation~\cite{Cui2019CPL} without introducing a structural phase transition.

%
%

To summarize, we carried out a detailed study on the optical properties of \BSTS, a TI with $E_F$ lying in the bulk band gap. The absence of Drude response in $\sigma_{1}(\omega)$ attests the bulk-insulating nature of this material. We observed a sharp peak at about 2\,200~cm$^{-1}$ in $\sigma_{1}(\omega)$ at 5~K. Further quantitative analyses of the line shape and $T$ dependence of the peak, in conjunction with first-principles calculations, unambiguously testify that the sharp peak in $\sigma_{1}(\omega)$ represents a VHS arising from Mexican-hat-shaped inverted bands.

%
%

\begin{acknowledgments}
We thank Bing Xu, Xiaoxiang Xi and Yuxin Zhao for illuminating discussions. We gratefully acknowledge financial support from the MOST of China (grant nos. 2016YFA0300400 and 2015CB921202), the National Natural Science Foundation of China (grant nos. 11874206, 11534005, 11374144, 11574133 and 11834006), and the Fundamental Research Funds for the Central Universities with grant no. 020414380095. The calculations were carried out using supercomputers at the High Performance Computing Center of Collaborative Innovation Center of Advanced Microstructures, the high-performance supercomputing center of Nanjing University and `Tianhe-2' at NSCC-Guangzhou.

Wenchao Jiang, Bowen Li and Xiaomeng Wang contributed equally to this work.
\end{acknowledgments}

%

\end{document}